\documentclass[pra,preprint,showpacs]{revtex4}
\usepackage[centertags]{amsmath}
\usepackage[utf8]{inputenc}
\usepackage{amsfonts}
\usepackage{amssymb}
\usepackage{amsthm}
\usepackage{newlfont}
\usepackage{epsfig}
\usepackage{amscd}
\usepackage{graphicx}
\usepackage{epsfig}
\usepackage{footnote}
\usepackage{lipsum}
\usepackage{color}
\usepackage{xcolor}
\usepackage{graphicx}
\usepackage{multirow}
\usepackage{enumerate}
\usepackage{caption}
\usepackage{subcaption}

\usepackage{amscd}

\newcommand{\beq}{\begin{equation}}
\newcommand{\eeq}{\end{equation}}
\newcommand{\ba}{\begin{array}}
\newcommand{\ea}{\end{array}}
\newcommand{\bea}{\begin{eqnarray}}
\newcommand{\eea}{\end{eqnarray}}
\begin{document}

\begin{center}
{\large \sc \bf {Transfer of 0-order coherence matrix along spin-1/2 chain}
}

\vskip 15pt

{\large
G.A.Bochkin$^{1,2}$, E.B.Fel'dman$^{1,2}$,  I.D.Lazarev$^{1,2}$, A.N.Pechen$^{2,3}$ and  A.I.~Zenchuk$^{1,2,*}$
}

\vskip 8pt

{\it $^1$Institute of Problems of Chemical Physics, RAS,
Chernogolovka, Moscow reg., 142432, Russia}.

{\it $^2$Steklov Mathematical Institute of Russian Academy of Sciences, Gubkina str. 8, Moscow 119991, Russia}

{\it $^3$National University of Science and Technology ''MISIS'', Leninski prosp. 4, Moscow 119049, Russia}

{\it $^*$Corresponding author. E-mail:  zenchuk@itp.ac.ru}
\vskip 8pt

\end{center}

\begin{abstract}
{In this work, we study transfer of coherence matrices along spin-1/2 chains of various length.} Unlike higher order coherence matrices, 0-order coherence matrix can be perfectly transferred if its elements are properly fixed. In certain cases, to provide the perfect transfer, an extended receiver together with optimized its unitary transformation has to be included into the protocol.{In this work,} the asymptotic  perfectly transferable 0-order coherence matrix for an infinitely long chain is considered and deviation of a perfectly transferred state from this asymptotic state is studied as a function of the chain length for various sizes of the extended receiver.
The problem of arbitrary parameter transfer via the nondiagonal elements of the 0-order coherence matrix  is also considered and optimized using the unitary transformation of the extended receiver.

\end{abstract}


\maketitle

\section{Introduction}
\label{Section:introduxtion}

{The methods of quantum state transfer cover an important area of quantum communication and demonstrate advantages of quantum mechanics approach in comparison with classical one. The problem of state transfer between different nodes  of a multiparticle quantum system is motivated not only by the needs of long-distance exchange of quantum information but also by the needs of exchanging quantum information among different  quantum devices transferring the output state of a particular circuit (sender) to the input register of another circuit (receiver). 
Similar problems appear in the general field of optimal quantum
control~\cite{arXiv:2205.12110}.
However, in general,  quantum evolution leads to spread of the initial quantum state of the sender over the whole system. Therefore, we need a mechanism which would lead to the collapse of the transferred state at the receiver. This motivates development of the models serving for either perfect or high-probability state transfer.}
The phenomenon of perfect state transfer (PST) was intensively investigated after the state transfer problem was formulated in Ref.~\cite{Bose}. It turned out that PST in spin chains is achievable in very specific cases \cite{CDEL,KS} and can  be easily destroyed by small perturbations of the interaction Hamiltonian.  In this case,  PST becomes high probability state transfer (HPST) \cite{ZASO}. HPST is simpler for realization \cite{GKMT,GMT} and  it demonstrates stability with respect to small perturbations of the Hamiltonian \cite{ZASO2}. 
{Of course, the privileged carriers for long-distance communications 
are photons \cite{PBGWK,PBGWK2,DLMRKBPVZBW}. However, spin-states can also serve for this purpose in compact quantum devices \cite{PSB,LH}.  This motivates our study in that direction.} 

One has to emphasize that the main idea of PST
is to use such coupling constants in the Hamiltonian that provide proper rational numbers for all ratios of eigenvalues of the Hamiltonian. {For instance, those coupling constants $D_{n}$ between the $n$th and  $(n+1)$th nodes   were found  to be $D_{n} = \lambda \sqrt{n( N-n)}$ ($\lambda$ is a normalization constant) in \cite{CDEL} for the $XX$ Hamiltonian with the nearest neighbor interaction.} Of course, this requirement is very sensitive to the variations in the environment, but nevertheless it serves a reference point in many state-transfer protocols.

{The high-probability state transfer uses different principle. While all eigenvalues of the Hamiltonian contribute to the state-transfer probability in the perfect state transfer, high-probability state transfer is based on selecting 
several 
eigenvalues of the interaction Hamiltonian (typically two or three) which yields the main contribution to the 
state-transfer probability. The most popular way to reach this aim is using the weak bonds between the end nodes and the main body of the chain  (the weak end-bond model) \cite{WLKGGB,GKMT,ZASO2}. However,  a specially adjusted local magnetic field can be also used  \cite{DZ_2010}. Application of these two methods can be found in Refs.\cite{PLAPG,FZ_2009,YB}. Many other aspects of state transfer process were considered in \cite{ABCVV,FR_2005,KF_2006,VGIZ,BK,JSTB,LPRA,HL,YB2}. 
}

Although HPST is more reliable then PST, the state transfer fidelity decreases with the chain length.
One of the methods to partially overcome this obstacle is to use the extended receiver (i.e., receiver joined with its few nearest nodes) together with a special unitary transformation, which was also effective in remote state preparation \cite{BZ_2018}.

Recently the concept of transfer of non-interacting multi-quantum coherence matrix was introduced  \cite{FZ_2017}.
Then, it was pointed in \cite{BFZ_Arch2018} that the zero-order coherence matrix of special form can be perfectly transferred  from the sender to the receiver along the  tripartite  spin system (which includes sender $S$, transmission line $TL$, and receiver $R$) with the only requirements that the Hamiltonian is conserving the excitation number of the spin system and, in addition, the  initial state of $TL\cup R$ must be a 0-order coherence matrix (it is a thermodynamic equilibrium state in \cite{BFZ_Arch2018}). Notice that the unitary transformation of the extended receiver was not used in \cite{BFZ_Arch2018}. Next, in \cite{FPZ_Arxiv2021}, general statements regarding the perfect transfer of a 0-order coherence matrix were formulated for the case of the ground initial state of the subsystem $TL\cup R$. In that case, an additional unitary transformation should be applied to the final receiver's state which exchanges two elements of the receiver density matrix: the elements corresponding to 0- and maximal excitation number. The important feature of that transformation is that it does not conserve the excitation number of the spin system.

Continuing the results of Refs.\cite{FZ_2017,BFZ_Arch2018,Z_2018,FPZ_Arxiv2021}, the concept of optimal state transfer was formulated.
This optimal state transfer is the structural restoring of the higher order coherence matrices of the initial sender's state and perfect transfer of the 0-order coherence matrix, or, if desired,  the structural restoring of the whole nondiagonal part of the receiver's initial state and perfect transfer of its diagonal part. Structural restoring of some blocks of the transferred density matrix means that each element of this block in the receiver's density matrix differs from the appropriate element of the sender's density matrix by a multiplicative factor. Such restoring is achievable due to using the optimizing unitary transformation of the extended receiver which plays a crucial role in the optimal state transfer. We emphasize that the above factors in the restored state as well as the optimizing transformation are universal objects which are defined by the interaction Hamiltonian and time instant for the receiver's state registration and they do not depend on the particular sender's initial state to be transferred.

All this motivates the detailed study of the 0-order coherence matrix which is the subject of our paper.
The basic problems to be explored are the following.
\begin{enumerate}
\item
As a preliminary step,  study the general block-structure of  density matrices  involved into the state-transfer process preserving the excitation number:
\begin{eqnarray}
&&
{\mbox{initial sender's density matrix $\rho^{(S)}(0)$}}  \;\; \to\;\;\\\nonumber
&& {\mbox{density matrix of the whole evolutionary system $\rho(t)$}}\;\;
\to\;\;
\\\nonumber
&&
{\mbox{receiver's density matrix at certain time instant $t_0$, $\rho^{(R)}(t_0)$}}.
\end{eqnarray}
\item
Determine the structure of the perfectly transferable 0-order coherence matrix (PTZ)  in an infinitely long chain (the asymptotic PTZ) and study deviation of the PTZ from the asymptotic one.
\item
{Explore and optimize the protocol for an arbitrary parameter transfer via encoding parameters in the elements of the 0-order coherence matrix.
Remark, that the possibility of encoding free parameters in the nondiagonal part of 0-order coherence matrix was shown in \cite{Z_2018}.  Here we show that arbitrary parameters can be encoded into all elements of one-excitation block of 0-order coherence matrix.}
\end{enumerate}

The paper is organized as follows. In Sec.~\ref{Section:dm} we discuss the general block structure of a density matrix of spin system and evolution of this structure. In Sec.~\ref{Section:0}, we study  the structure of 0-order coherence matrix and  the perfect transfer of this matrix from the sender to the receiver. The asymptotic perfectly transferable state for infinitely long chain
is proposed, and the difference between the {norms of the  asymptotic state and the perfectly transferred state} is studied as a function of the chain length. A particular case of the  perfect transfer of the 0-order coherence matrix including only two blocks corresponding to  0- and 1-excitation is explored in Sec.~\ref{Section:01}. {The arbitrary parameter transfer using the elements of the one-excitation block  is also considered in that section.} Conclusions are provided in Sec.~\ref{Section:conclusion}. Some important details regarding the structure of the transfer matrix (i.e., of the matrix which transfers the initial sender state to the final receiver state) are collected in Appendix, Sec.~\ref{Section:A}.

\section{General structure of density matrix}
\label{Section:dm}

Any density matrix $\rho$ of a spin chain can be represented as the sum of multi-quantum (MQ) coherence matrices:
\begin{eqnarray}\label{rhon}
\rho=\sum_{n=-N}^N \rho^{(n)},\quad \textrm{ such that}\quad [I_z,\rho^{(n)}]=n \rho^{(n)},
\end{eqnarray}
where $\rho^{(n)}$ is the $n$-order coherence matrix, $I_z=\sum_i I_{zi}$ is the $z$-projection of the total spin of the chain, $I_{zi}$ is the $z$-projection of the $i$th spin.
Notice that the matrix  $\rho^{(n)}$  in (\ref{rhon}) collects  the probability amplitudes of state transitions  which increase (for $n>0$) or decrease (for $n<0$) the  excitation number of states by $n$. There are $2N+1$ such matrices in the sum.

According to the definition of  $\rho^{(n)}$, the $N$-qubit density matrix $\rho$, has the following block-structure:
\begin{eqnarray}\label{blocks}
\rho= \left(\begin{array}{c|c|c|c|c|c}
\sigma^{(0)}_{0,0}&\sigma^{(1)}_{0,1}&\sigma^{(2)}_{0,2}&\cdots& \sigma^{(N-1)}_{0,N-1}&\sigma^{(N)}_{0,N}\cr\hline
\sigma^{(-1)}_{1,0}&\sigma^{(0)}_{1,1}&\sigma^{(1)}_{1,2}&\cdots& \sigma^{(N-2)}_{1,N-1}&\sigma^{(N-1)}_{1,N}\cr\hline
\sigma^{(-2)}_{2,0}&\sigma^{(-1)}_{2,1}&\sigma^{(0)}_{2,2}&\cdots& \sigma^{(N-3)}_{2,N-1}&\sigma^{(N-2)}_{2,N}\cr\hline
\cdots&\cdots&\cdots&\cdots&\cdots&\cdots\cr\hline
\sigma^{(-N+1)}_{N-1,0}&\sigma^{(-N+2)}_{N-1,1}&\sigma^{(-N+3)}_{N-1,2}&\cdots& \sigma^{(0)}_{N-1,N-1}&\sigma^{(1)}_{N-1,N}\cr\hline
\sigma^{(-N)}_{N,0}&\sigma^{(-N+1)}_{N,1}&\sigma^{(-N+2)}_{N,2}&\cdots& \sigma^{(-1)}_{N,N-1}& \sigma^{(0)}_{N,N}\cr
\end{array}\right),\qquad \sigma^{(-k)}_{ij}=(\sigma^{(k)}_{ji})^\dagger  .
\end{eqnarray}
Each block $\sigma^{(n)}_{i,j}$ is included into the $n$-order coherence matrix and has dimension $D_{i}\times D_{j}$ ($n=j-i$),
where $D_{i}$ is the dimension of the $i$-excitation subspace which, for the $N$-qubit system, reads:
\begin{eqnarray}
\; D_{k}=
\left(\begin{array}{c}
N\cr k \end{array}\right) ,\;\;k=0,1,\dots,N.
\end{eqnarray}
In particular, $D_{0}=D_{N}=1$, $D_{1}=D_{N-1}=N$. Therefore, the blocks in (\ref{blocks}) are in general not square matrices.

\subsection{Evolution conserving excitation number}
We consider the evolution of the spin chain governed by the  $XX$-Hamiltonian
\begin{eqnarray}\label{XXall}\label{XX}
H=\sum_{j>i} D_{ij}(I_{i;x}I_{j;x}+I_{i;y}I_{j;y}),
\end{eqnarray}
where $D_{ij}=\gamma^2 \hbar/r_{ij}^3$ is the coupling constant between the $i$th and $j$th spins (the magnetic field is directed along the chain),  $\gamma$ is the gyromagnetic ratio, and
$\hbar$  is the Planck constant. For the homogeneous chain we have $r_{i,i+1}=r$ and therefore the coupling constants between the nearest neighbors are the same. Hereafter we consider the homogeneous spin chain.
Hamiltonian (\ref{XX}) satisfies the commutation relation
\begin{eqnarray}\label{comm}
[H,I_z]=0,
\end{eqnarray}
and therefore has the following block-diagonal structure:
\begin{eqnarray}\label{block}
H={\mbox{diag}}(H^{(0)},H^{(1)},\dots,H^{(N)}),
\end{eqnarray}
where the dimension of the block $H^{(n)}$ is $D_n\times D_n$.  In this formula, the $n$th block governs the evolution of the
subspace of the $n$-excitation states, and there are $N+1$ such blocks. The evolution operator $V(t)=e^{-i H t}$ generated by the Hamiltonian (\ref{block}) also has the  block-diagonal structure
\begin{eqnarray}\label{bV}
V={\mbox{diag}}(V^{(0)},V^{(1)},\dots,V^{(N)}),\;\;V^{(n)} = e^{-i H^{(n)} t}
\end{eqnarray}
and the  dimension of the block  $V^{(n)}$ is the same as the dimension of $H^{(n)}$.
This means that each block $\sigma^{(n)}_{i,j}$ in (\ref{blocks})  evolves independently:
\begin{eqnarray}
\label{b_ev}
\rho(t) = V(t) \rho(0) V^\dagger(t)\;\;\Rightarrow \;\;
\sigma^{(n)}_{i,j}(t) = V^{(i)}(t) \sigma^{(n)}_{i,j}(0)  (V^{(j)}(t))^+.
\end{eqnarray}

\subsection{State transfer along  spin chain}
Now we consider the communication line including the sender $S$, transmission line $TL$ and receiver $R$ and describe the state propagation from the sender to the receiver.

Suppose that we have $K$-excitation initial state of the $N^{(S)}$-qubit sender, $K\le N^{(S)}$:
\begin{eqnarray}\label{blocksS}
\rho^{(S)}= \left(\begin{array}{c|c|c|c|c|c}
s^{(0)}_{0,0}&s^{(1)}_{0,1}&s^{(2)}_{0,2}&\cdots& s^{(K-1)}_{0,K-1}&s^{(K)}_{0,K}\cr\hline
s^{(-1)}_{1,0}&s^{(0)}_{1,1}&s^{(1)}_{1,2}&\cdots& s^{(K-2)}_{1,K-1}&s^{(K-1)}_{1,K}\cr\hline
s^{(-2)}_{2,0}&s^{(-1)}_{2,1}&s^{(0)}_{2,2}&\cdots& s^{(K-3)}_{2,K-1}&s^{(K-2)}_{2,K}\cr\hline
\cdots&\cdots&\cdots&\cdots&\cdots&\cdots\cr\hline
s^{(-K+1)}_{K-1,0}&s^{(-K+2)}_{K-1,1}&s^{(-K+3)}_{K-1,2}&\cdots& s^{(0)}_{K-1,K-1}&s^{(1)}_{K-1,K}\cr\hline
s^{(-K)}_{K,0}&s^{(-K+1)}_{K,1}&s^{(-K+2)}_{K,2}&\cdots& s^{(-1)}_{K,K-1}& s^{(0)}_{K,K}\cr
\end{array}\right).
\end{eqnarray}
The dimension of each block $s^{(k)}_{i,j}$  is $D^{(S)}_i \times D^{(S)}_j$,
\begin{eqnarray}
\; D^{(S)}_{k}=
\left(\begin{array}{c}
N^{(S)}\cr k \end{array}\right) ,\;\;k=0,1,\dots,K.
\end{eqnarray}
As was shown in \cite{FZ_2017}, to arrange the independent propagation of the MQ-coherence matrices from the sender to the receiver, two following sufficient conditions  must be imposed on the initial state.
\begin{enumerate}
\item
The initial state  should have tensor-product form:
\begin{eqnarray}
\rho(0)=\rho^{(S)}(0)\otimes \rho^{(TL,R)}(0),
\end{eqnarray}
where $\rho^{(S)}(0)$ and $\rho^{(TL,R)}(0)$ are, respectively,  the initial states of the sender and transmission line joined with the receiver.
\item
The initial state $\rho^{(TL,R)}(0)$ should include only 0-order coherence matrix; it is the ground state in our paper.
\end{enumerate}
Then the evolution of the state of the whole $N$-qubit  system {$S\cup TL\cup R$} is described by the
 density matrix $\rho$ having the block-structure (\ref{blocks}).
We emphasize that the excitation number remains the same, and the elements of the particular block $s^{(n)}_{i,j}$ appear only in the appropriate block $\sigma^{(n)}_{i,j}$, i.e., {we have  the following
 block-map}:
\begin{eqnarray}\label{S_L}
s^{(k)}_{ij} \to \sigma^{(k)}_{ij},\;\;k=1,\dots,K,
\end{eqnarray}
{where dimensions of $\sigma$-blocks are larger than dimensions of $s$-blocks.}

Next, the state of the receiver at some time instant $t$ reads
\begin{eqnarray}\label{blocksR}
\rho^{(R)}={\mbox{Tr}}_{S,TL}\rho(t)=\left(
\begin{array}{c|c|c|c|c|c}
r^{(0)}_{0,0}&r^{(1)}_{0,1}&r^{(2)}_{0,2}&\cdots& r^{(K-1)}_{0,K-1}&r^{(K)}_{0,K}\cr\hline
r^{(-1)}_{1,0}&r^{(0)}_{1,1}&r^{(1)}_{1,2}&\cdots& r^{(K-2)}_{1,K-1}&r^{(K-1)}_{1,K}\cr\hline
r^{(-2)}_{2,0}&r^{(-1)}_{2,1}&r^{(0)}_{2,2}&\cdots& r^{(K-3)}_{2,K-1}&r^{(K-2)}_{2,K}\cr\hline
\cdots&\cdots&\cdots&\cdots&\cdots&\cdots\cr\hline
r^{(-K+1)}_{K-1,0}&r^{(-K+2)}_{K-1,1}&r^{(-K+3)}_{K-1,2}&\cdots& r^{(0)}_{K-1,K-1}&r^{(1)}_{K-1,K}\cr\hline
r^{(-K)}_{K,0}&r^{(-K+1)}_{K,1}&r^{(-K+2)}_{K,2}&\cdots& r^{(-1)}_{K,K-1}& r^{(0)}_{K,K}\cr
\end{array}\right),
\end{eqnarray}
i.e., we have another map
\begin{eqnarray}\label{sigmar}
\sigma^{(k)}_{ij} \to r^{(k)}_{ij}.
\end{eqnarray}
Again,  matrix (\ref{blocksR}) includes up to $K$-excitation blocks and map (\ref{sigmar}) decreases the dimension of each block   reducing it from $D_i\times D_j$ to  $D^{(R)}_i \times D^{(R)}_j$, $D^{(R)}_i<D_i$,
\begin{eqnarray}
\; D^{(R)}_{k}=
\left(\begin{array}{c}
N^{(R)}\cr k \end{array}\right) ,\;\;k=0,1,\dots,K.
\end{eqnarray}
In particular, if $N^{(S)}=N^{(R)}$, then $D^{(R)}_i = D^{(S)}_i$.
Notice that calculating ${\mbox{Tr}}_{TL,S} \;\rho$  we calculate the trace of each block $\sigma^{(k)}_{i,j}$. Since $j=i+k$ in the notation $\sigma^{(k)}_{i,j}$ and the coherence order of each particular block is conserved by  trace operation \cite{FZ_2017}, we can write
\begin{eqnarray}\label{red0}
{\mbox{Tr}}_{S,TL}\sigma^{(k)}_{i,i+k} = \sum_{l=0}^i \tilde\sigma^{(k;i)}_{l,l+k},\;\;i=0,\dots,K-k,
\end{eqnarray}
where each term $\tilde\sigma^{(k;i)}_{l,l+k}$ contributes into the block  $r^{(k)}_{l,l+k}$.
Therefore we have
\begin{eqnarray}\label{red}
r^{(k)}_{l,l+k} = \sum_{i=0}^{K-k}  \tilde \sigma^{(k,i)}_{l,l+k}.
\end{eqnarray}

Hereafter we concentrate on the evolution of the 0-order coherence matrix. We show that the perfectly  transferable 0-order coherence matrix   can be constructed for  the case $K=N^{(S)}$ without involving extended receiver, while for the case $K<N^{(S)}$  we have to involve special unitary transformation of the extended receiver to reach the goal.

\section{Transfer of 0-order coherence matrix}
\label{Section:0}
Hereafter we study the 0-order coherence matrix and adopt the following notation:
\begin{eqnarray}\label{zeroorder}
&&
s^{(0)}_{k,k}=s^{(k)},\\\nonumber
&&
r^{(0)}_{k,k}=r^{(k)},\\\nonumber
&&
\sigma^{(0)}_{k,k}=\sigma^{(k)}.
\end{eqnarray}
Therefore, below the subscripts mean a particular elements of the appropriate block, for instance $s^{(k)}_{ij}$.
In the case of the sender  initial state including only 0-order coherence matrix formulae (\ref{blocksS}), (\ref{blocks}), (\ref{blocksR}) read, respectively,
\begin{eqnarray}\label{blocksS0}
&&
\rho^{(S)} = {\mbox{diag}} (s^{(0)},\dots, s^{(K)}),\\
\label{blocksSTLR0}
&&
\rho = {\mbox{diag}} (\sigma^{(0)},\dots, \sigma^{(K)}),\\
\label{blocksR0}
&&
\rho^{(R)} = {\mbox{diag}} (r^{(0)},\dots, r^{(K)}).
\end{eqnarray}
and formulas (\ref{red0}) reduce to
\begin{eqnarray}\label{red02}
{\mbox{Tr}}_{S,TL}\sigma^{(i)} = \sum_{l=0}^i \tilde\sigma^{(i;l)},\;\;i=0,\dots,K,
\end{eqnarray}
where each term $\tilde\sigma^{(i;l)}$ contributes to the block  $r^{(l)}$,
and (\ref{red})
gets the form
\begin{eqnarray}\label{red2}
r^{(l)} = \sum_{i=l}^{K}  \tilde \sigma^{(i;l)}.
\end{eqnarray}
Thus,
 for a fixed $l$, each block $r^{(l)}$ depends not only on the elements of the block
$\sigma^{(l)}$, but on the elements of the blocks $\sigma^{(j)}$ with $j>l$. In particular, the block $r^{(N^{(R)})}$ has no contributions from any $\sigma$-block, while the block  $r^{(0)}$ includes contributions from all $\sigma$-blocks.

\subsubsection{Asymptotic receiver's state as $N\to\infty$ }

The asymptotic receiver's state {$\rho^{(R;0)}_\infty$ for infinitely long chain} is prompt by  formula (\ref{red2}) and by the fact that {all elements $\rho_{ij}$ vanish with an increase in the chain length} for a fixed sender's dimension $N^{(S)}$. In other words, the asymptotic state of the receiver is expected to be the ground state
{ because the 0-excitation block of $\rho^{(R)}$ gathers elements from the higher-excitation  blocks due to the partial trace}. Thus
\begin{eqnarray}\label{rhoRas}
\rho^{(R;0)}_{\infty} = {\mbox{diag}}(1,0,\dots,0).
\end{eqnarray}

\subsection{Unitary transformation of the extended receiver}
To handle the structure of the nondiagonal elements of the zero-order coherence matrix, we use the unitary transformation $U$ of the so-called extended receiver (the receiver joined with its several neighboring spins).
The unitary transformation of the $N^{(ER)}$-qubit  extended receiver has also the diagonal block structure:
\begin{eqnarray}\label{U}
U(\varphi)={\mbox{diag}}(1,U^{(1)}(\varphi^{(1)}) ,\dots,U^{(N^{(ER)}}(\varphi^{(N^{(ER)}})),
\end{eqnarray}
where $\varphi=(\varphi^{(1)},\dots,\varphi^{(N^{(ER)}})$, and $\varphi^{(k)}$ are the sets of free parameters in the block $U^{(k)}$:
$\varphi^{(k)}= (\varphi^{(k)}_1,\dots, \varphi^{(k)}_{F^{(k)}})$. Here $F^{(k)}$ is the
parameter defined below in Eq.~(\ref{utr}).

The parametrization of a particular block $U^{(k)}$ can be done as follows. Let us enumerate the nondiagonal  elements of the upper triangular submatrix of a $D_k\times D_k$ matrix as follows. The nondiagonal element in the $k$th row and $l$th column, $l>k$, prescribes the index $n$,
\begin{eqnarray}\label{n}
n=\sum_{m=1}^{k-1}(D_k - m) + l-k.
\end{eqnarray}
Then the block $U^{(k)}$ of the unitary transformation can be parameterized as follows:
\begin{eqnarray}
U^{(k)} =\prod_{n=1}^{F^{(k)}} e^{i \sigma^{(k)}_{xn} \varphi^{(k)}_n}
\prod_{n=1}^{F^{(k)}} e^{i \sigma^{(k)}_{yn} \varphi^{(k)}_n}.
\end{eqnarray}
Here $\sigma^{(k)}_{xn}$ with $n$ defined in (\ref{n}) is the matrix  with two units in the $k$th row and $l$th column and in the
$l$th row and $k$th column. Similarly,
$\sigma^{(k)}_{yn}$ with $n$ defined in (\ref{n}) is the matrix  with $-i$  ($i^2=-1$) in the $k$th row and $l$th column and $i$ in the
$l$th row and $k$th column.

As was stated in Ref.~\cite{Z_2018}, only nondiagonal elements of $U$ are effective.
Therefore the number  of free real non-diagonal  parameters in the $k$th block of $U$ is
\begin{eqnarray}\label{utr}
F^{(k)} = D^{(ER)}_{k}(D^{(ER)}_{k}-1), \;\;k=0,\dots, N^{(ER)}.
\end{eqnarray}
Thus, the total number of free parameters is (since $F^{(0)}=F^{(N)}=0$) is
\begin{eqnarray}
F = \sum_{n=1}^{N-1} F^{(n)}.
\end{eqnarray}
However, only blocks of $U$  with up to $K$ excitations are effective, where $K$ is the excitation number in
$\rho^{(S)}(0)$. Therefore, the number of effective parameters is
\begin{eqnarray}
F_{eff} = \sum_{n=1}^{K} F^{(n)}.
\end{eqnarray}
We combine the  evolution operator $V$ and transformation $U$ into the single operator
$W$:
\begin{eqnarray}\label{WW}
W(t)=\Big(\mathbb I_{S,TL'}\otimes U\Big) V(t),
\end{eqnarray}
where $TL'$ is the transmission line without the nodes of the extended receiver. Of course, $W$ has the block-structure similar to (\ref{bV}):
\begin{eqnarray}\label{bW}
W={\mbox{diag}}(W^{(0)},W^{(1)},\dots,W^{(N)})
\end{eqnarray}
with scalar blocks $W^{(0)}$ and $W^{(N)}$.

Although using the unitary transformation we can not completely restore the diagonal elements of the 0-order coherence matrix, but we can use these parameters to
restore the nondiagonal elements of the 0-order coherence matrix. This can be useful, in particular, to keep the required form  of the zero-order coherence matrix, as will be used below.

\subsection{Perfect transfer of the 0-order coherence matrix}

\subsubsection{Complete state space of the sender}
\label{Section:complete}
First we consider $N^{(S)}$-excitation initial state of the $N^{(S)}$-qubit  sender  and transfer this state to the $N^{(R)}=N^{(S)}$-qubit receiver, i.e., the transferred  density matrix  consists of all blocks
related with the excitation numbers  from 0 to $N^{(S)}$
(the complete state space of sender).

\begin{figure*}[!]
\epsfig{file=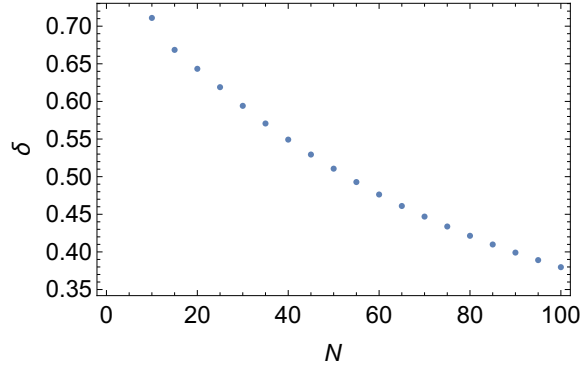,
  scale=0.6
   ,angle=0
}
\caption{Deviation $\delta$ of the perfectly transferred 2-qubit state $\rho^{(S)}$ ($N^{(S)}=2$) from the asymptotic density matrix
$\rho^{(S)}_{\infty}$.
}
  \label{Fig:F}
\end{figure*}

In this case, the sender and receiver  states are
\begin{eqnarray}\label{compl}
\rho^{(S)}  ={\mbox{diag}}(s^{(0)},s^{(1)},\dots,s^{(N^{(S)})}), \;\;\;
\rho^{(R)}(t)  ={\mbox{diag}}(r^{(0)}(t),r^{(1)}(t),\dots,r^{(N^{(S)})}(t)),
\end{eqnarray}
where  $s^{(0)}$,  $s^{(N^{(S)})}$, $r^{(0)}$ and $r^{(N^{(S)})}$ are scalars with
$r^{(N)}(t) = |W^{(N)}(t)|^2 s^{(N)}$.
{We select the time instant for state registration $t^{(N^{(S)})}$ corresponding to the maximum of $|W^{(N)}(t)|^2$ (the probability of the $N^{(S)}$-excitation state transfer from the sender to the receiver \cite{FPZ_Arxiv2021}):
\begin{eqnarray}
\max_t \left(|W^{(N)}(t)|^2 \right) =|W^{(N)}(t^{(N^{(S)})})|^2.
\end{eqnarray}
This time instant  is almost a linear function of the chain length $N$, as is demonstrated in
Table \ref{Table:time} for the $XX$-Hamiltonian (\ref{XX}) with $N^{(S)}=2$, see also Fig.\ref{fig:time-instant-n}.

For a long homogeneous chain one has $|W^{(N^{(S)})}|^2<1$, and
therefore the equality $\rho^{(R)}(t)=\rho^{(S)}(0)$ is impossible. But we can find such $s^{(0)}(0)$ that (see~\cite{FPZ_Arxiv2021})
\begin{eqnarray}\label{syst12}
r^{(N^{(S)})}(t^{(N^{(S)})}) = s^{(0)}(0).
\end{eqnarray}
We also require the elements of $s^{(k)}(0)$ ($k\neq 0$, $N^{(S)}$) to satisfy the following equations
at the time instant $t^{(N^{(S)})}$:
\begin{eqnarray}\label{syst1}
&&r^{(k)}(t^{(N^{(S)})}) = s^{(k)}(0),\;\;1\le k\le N^{(S)}-1.
\end{eqnarray}}
Then $\rho^{(R)}(t^{(N^{(S)})})$ coincides with $\rho^{(S)}(0)$ up to the exchange of two elements of blocks with zero and $N^{(S)}$ excitations.
This exchange can be performed by the unitary transformation $U^{ex}$ such that
\begin{eqnarray}\label{comm2}
[U^{ex},I_z]\neq 0.
\end{eqnarray}
%
\begin{table}
\begin{tabular}{|c|cccccccccc|}\hline
 N & 10 & 15 & 20 & 25 & 30 & 35 & 40 & 45 & 50 & 55 \\\hline
 $t^{(2)}$&12.8896 & 18.2026 & 23.4171 & 28.5937 & 33.7448 & 38.8736 & 43.9842 & 49.0820 &
   54.1709 & 59.2527 \\\hline
\end{tabular}
\begin{tabular}{|c|ccccccccc|}\hline
 $N$ & 60& 65 & 70 & 75 & 80 & 85 & 90 & 95 & 100 \\\hline
$t^{(2)}$& 64.3271 & 69.3941 & 74.4548 & 79.5106 & 84.5620 & 89.6089 & 94.6516 & 99.6898 &
   104.724 \\\hline
\end{tabular}
\caption{Time instants for state registration for spin chain of different chain length $N$ governed by $XX$-Hamiltonian (\ref{XX}) with  2-qubit sender including up to two excitations.}
\label{Table:time}
\end{table}
An interesting question is the dependence of the elements of the  perfectly transferred zero-order coherence  matrix on the chain length $N$. { The asymptotic  receiver's state (\ref{rhoRas}) allows to assume that the sender state found as a solution of the system (\ref{syst12}), (\ref{syst1}) in view of the unitary transformation  $U^{ex}$  tends to
\begin{eqnarray}\label{limit}
\rho^{(R)}_{\infty}=\rho^{(S)}_{\infty}= {\mbox{diag}}(0,0,\dots, 0,1)
\end{eqnarray}
in a long chain.}
The graph of the deviation
\begin{eqnarray}\label{del}
\delta = ||\rho^{(S)}_\infty- \rho^{(S)}(N)||, 
\end{eqnarray}
where $||A|| =\sqrt{{\mbox{Tr}} (A A^\dagger)}$ is the Frobenius (or the Hilbert–Schmidt) norm, as a function of $N$ for the 2-qubit sender at time instants taken from Table~\ref{Table:time} is shown in Fig.~\ref{Fig:F}. In this case we have three blocks in the sender and receiver density matrices.
\paragraph{Effect of PTZ on the structure of the restorable higher-order coherence matrices.}
We emphasize that the unitary transformation $U^{ex}$ interchanges the first column (respectively, first row) with the last column (respectively, last row) of the matrix $\rho^{(R)}$. {Therefore, using the proposed protocol for the perfect transfer of the  0-order coherence matrix in combination with   the structural restoring of the higher order coherence matrices imposes a restriction on the structure of the optimally transferable sender's state \cite{FPZ_Arxiv2021}.}
Namely, the sender's density matrix (\ref{blocksS}) now should have the following form (using notation (\ref{zeroorder}) for the diagonal blocks):
\begin{eqnarray}\label{blocksS2}
\rho^{(S)}=\left( \begin{array}{c|c|c|c|c|c}
 s^{(0)}&0^{(1)}_{0,1}&0^{(2)}_{0,2}&\cdots& 0^{(N^{(S)}-1)}_{0,N^{(S)}-1} &0^{(N^{(S)})}_{0,N^{(S)}} \cr\hline
0^{-1}_{1,0}&s^{(1)}&s^{(1)}_{1,2}&\cdots& s^{(N^{(S)}-2)}_{1,N^{(S)}-1}&0^{(N^{(S)}-1)}_{1,N^{(S)}}
\cr\hline
0^{(-2)}_{2,0}&s^{(-1)}_{2,1}&s^{(2)}&\cdots& s^{(N^{(S)}-3)}_{2,N^{(S)}-1}& 0^{(N^{(S)}-2)}_{2,N^{(S)}}   \cr\hline
\cdots&\cdots&\cdots&\cdots&\cdots&\cdots\cr\hline
0^{(-N^{(S)}+1)}_{N^{(S)}-1,0} &s^{(-N^{(S)}+2)}_{N^{(S)}-1,1}&s^{(-N^{(S)}+3)}_{N^{(S)}-1,2}&\cdots& s^{(N^{(S)}-1)}&0^{(1)}_{N^{(S)}-1,N^{(S)}}
\cr\hline
0^{(-N^{(S)})}_{N^{(S)},0} &0^{(-N^{(S)}+1)}_{N^{(S)},1} &0^{(-N^{(S)}+2)}_{N^{(S)},2}
&\cdots&0^{(-1)}_{N^{(S)},N^{(S)}-1}& s^{(N^{(S)})}\cr
\end{array}\right),
\end{eqnarray}
{ where $0^{(n)}_{i,j}$ is  the block $s^{(n)}_{i,j}$ with all zeros}. The diagonal blocks in (\ref{blocksS2}) correspond to 0-order coherence matrix; for the blocks of non-zero order coherence matrices we keep the notations introduced in (\ref{blocksS}).
Formula (\ref{blocksS2}) shows that the first column (respectively, row) and the last column (respectively, row)  of the whole matrix
$\rho^{(S)}(0)$ must be zero except of their diagonal elements. Otherwise, the final unitary transformation $U^{ex}$ will mix some elements of coherence matrices of different orders.

\subsubsection{Restricted state space of the initial sender's state}
Suppose that the state of the  $N^{(S)}$-qubit sender includes $K<N^{(S)}$ excitations. Then, instead of (\ref{compl}) we have
\begin{eqnarray}\label{compl2}
\rho^{(S)}  ={\mbox{diag}}(s^{(0)},s^{(1)},\dots,s^{(K)}),\;\;
\rho^{(R)}  ={\mbox{diag}}(r^{(0)},r^{(1)},\dots,r^{(K)}).
\end{eqnarray}
The state in the form~(\ref{compl2}) can not be perfectly transferred using the method in Sec.~\ref{Section:complete}.  	In fact, this method assumes that the final unitary transformation $U^{ex}$ exchanges the positions of $r^{(0)}$ and one more  diagonal element, say  $r^{(K)}_{11}$. But such transformation does not conserve the excitation number (see Eq.~(\ref{comm2})), it   unavoidably exchanges rows and columns associated with the two mentioned diagonal  elements and thus creates  higher-order coherence matrices which are not desirable. To avoid this effect, we have to impose a special restriction on the structure of the blocks $s^{(K)}$ and $r^{(K)}$. Namely, let all the elements in the row and the column of the diagonal element  $s^{(K)}_{11}$  (i.e., all the elements of the first column and first row of the $K$th block) be zeros except the single diagonal element $s^{(K)}_{1,1}$, i.e.
\begin{eqnarray}\label{ssK}
s^{(K)}_{i,1}=s^{(K)}_{1,i} = 0,\;\;1<i\le D^{(S)}_K.
\end{eqnarray}
Then  the structure of $s^{(K)}$ reduces to the following one:
\begin{eqnarray}\label{tssK}
 s^{(K)}=
\left(\begin{array}{c|c}
s^{(K)}_{1,1}&0_{1,D^{(S)}_K-1}\cr
\hline
0_{D^{(S)}_K-1,1}&\tilde s^{(K)}\cr
\end{array}\right),
\end{eqnarray}
and
\begin{eqnarray}\label{ts}
\tilde s^{(K)}=\left(
\begin{array}{ccc}
s^{(K)}_{2,2}&\dots&s^{(K)}_{2,D^{(S)}_K}\cr
\dots&\dots&\dots\cr
(s^{(K)}_{2,D^{(S)}_K})^*&\dots&s^{(K)}_{ D^{(S)}_K,D^{(S)}_K}
\end{array}
\right),
\end{eqnarray}
{
where $0_{i,j}$ is the $i\times j$ zero matrix.
Recall that the subscripts in (\ref{ssK})-(\ref{ts}) are indexes of matrix elements.}

Now we can replace the system (\ref{syst1}) with the following one:
\begin{eqnarray}\label{rsred21}
&&r^{(k)} =s^{(k)},\;\;1\le k\le K-1,\\\label{rsred22}
&&r^{(K)}_{i,j} = s^{(K)}_{i,j}, \;\; 2\le i \le D^{(S)}_K,
\;\;2\le i \le D^{(S)}_K
,\\\label{rsred3}
&& s^{(K)}_{1,i} =0, \;\; 2\le i \le D^{(S)}_K,\\\label{rsred4}
&&r^{(K)}_{1 ,1} =s^{(0)},\\\label{rsred5}
&& r^{(K)}_{1,i} =0, \;\; 2\le i \le D^{(S)}_K
.
\end{eqnarray}

However, in the receiver density matrix $\rho^{(R)}$,
the elements  $ r^{(K)}_{i ,1}$ are non-zero  at $t>0$
because of  the mixing of the elements during the evolution. {That is why Eq.~(\ref{rsred5}) is valuable.}
Therefore, we have to involve the unitary transformation of the extended receiver with parameters $\varphi$ to make these elements zero at certain time instant $t^{(N^{(S)})}_K$ (this time instant will be specified below).
To write the system for the set of  parameters $\varphi$ of the unitary transformation, we use the results of \cite{FPZ_Arxiv2021} and write the relation between $r^{(k)}$ and $s^{(k)}$ as
\begin{eqnarray}\label{rT}
r^{(k)}_{i,j}(t,\varphi)=\sum_{n,m=1}^{D^{(S)}_{k}}T^{(k)}_{ij;nm}(t,\varphi) s^{(k)}_{nm}(0),\;\;
T^{(k)}_{ij;nm}(t,\varphi)=(T^{(k)}_{ji;mn}(t,\varphi))^*.
\end{eqnarray}
where $T^{(k)}$ is the transfer matrix of the $k$-excitation block whose explicit form is not important at the moment.
{
Then system (\ref{rsred5}) takes the form
\begin{eqnarray}\label{rT2}
r^{(k)}_{1,i}(t,\varphi)=\sum_{n,m=1}^{D^{(S)}_{k}}T^{(k)}_{1i;nm}(t,\varphi) s^{(k)}_{nm}(0) =0.
\end{eqnarray}
This system of equations is satisfied if
the parameters $\varphi$ at $t=t^{(N^{(S)})}_K$ satisfy the following system (taking into account (\ref{rsred3})):}
\begin{eqnarray}\label{T}
&&T^{(K)}_{1,n;i,i}(t^{(N^{(S)}_K)},\varphi)=0,\;\;2\le n\le D^{(S)}_K \;\;1\le i\le D^{(S)}_K,\\\label{T2}
&&{T^{(K)}_{1,n;i,j}(t^{(N^{(S)})}_K,\varphi)=0},\;\;2\le n\le D^{(S)}_K,\;\;
2\le i \le D^{(S)}_K,\;\;
2\le j \le D^{(S)}_K,\;\;i\neq j.
\end{eqnarray} 
Notice, that the asymptotic PTZ has the form (\ref{limit}) as well.

{{\bf Remark.} In principle, the system (\ref{rsred21})--(\ref{rsred5}) can be solved as the whole for the parameters $\varphi$ and elements of $r^{(k)}$. But we select Eq.~(\ref{rsred5}), replace it by system (\ref{T}), (\ref{T2}) and solve the later for $\varphi$. After that, the system (\ref{rsred21})--(\ref{rsred4}) can be solved for the elements of $r^{(k)}$.}

\section{0-order coherence matrix with up to one excitation}
\label{Section:01}
\subsection{Perfect transfer of 0-order coherence matrix}
\label{Section:01p}
We consider the 0-order coherence matrix including 0- and 1-excitation.
{According to (\ref{blocksS2}) and (\ref{tssK}), the initial sender's density matrix has the following structure:
\begin{eqnarray}\label{BlocksS3}
\rho^{(S)}=\left(
\begin{array}{c|c}
s^{(0)} & 0^{(1)}_{0,1}\cr
\hline
0^{-1}_{1,0}& s^{(1)}
\end{array}
\right),\;\; 
 s^{(1)}=
\left(\begin{array}{c|c}
s^{(1)}_{1,1}&0_{1,N^{(S)}-1}\cr
\hline
0_{N^{(S)}-1,1}&\tilde s^{(1)}\cr
\end{array}\right),
\end{eqnarray}
where $\tilde s^{(1)}$ is a full matrix of the form (\ref{ts}).
Since the only higher order coherence matrices  in this case are the $\pm 1$-order coherence matrices and they are zeros according to (\ref{BlocksS3}), this case doesn't allow to transfer any arbitrary parameter through the higher order coherence matrices. Nevertheless, we consider this case in details to reveal some general features of PTZ.
}

In this case the system (\ref{rT}) can be written in the simple matrix form
\begin{eqnarray}\label{calW}
r^{(1)}(t,\varphi)={\cal{W}}(t,\varphi) s^{(1)} {\cal{W}}^\dagger(t,\varphi),
\end{eqnarray}
see Appendix \ref{Section:A} for more detail.
Then system (\ref{T}), (\ref{T2})  with $K=1$ and $D^{(S)}_1=N^{(S)}$ reads
\begin{eqnarray}\label{T1}
T^{(1)}_{1,n;i,i}(t^{(N^{(S)})}_1,\varphi)&=&
{\cal{W}}_{1i}(t^{(N^{(S)})}_1,\varphi) {\cal{W}}^*_{ni}
(t^{(N^{(S)})}_1,\varphi)=0,\\\nonumber&&
2\le n\le N^{(S)}, \;\;1\le i\le N^{(S)},\\\label{T12}
T^{(1)}_{1,n;i,j}(t^{(N^{(S)})}_1,\varphi)&=&{\cal{W}}_{1i}(t^{(N^{(S)})}_1,\varphi) {\cal{W}}^*_{nj}
(t^{(N^{(S)})}_1,\varphi)0,\\\nonumber&&2\le n\le N^{(S)},\;\;
2\le i \le N^{(S)},\;\;
2\le j \le N^{(S)},\;\;i\neq j.
\end{eqnarray} 
This system is satisfied if $\varphi$ solves the smaller system of $N^{(S)}$ equations
\begin{eqnarray}\label{TW}
{\cal{W}}_{1j}=0,\;\;j=1,\dots, N^{(S)},
\end{eqnarray}
although solution space of   system (\ref{T1}), (\ref{T12}) is richer then that of (\ref{TW}).
Notice also that solution of  (\ref{TW}) leads to
\begin{eqnarray}\label{additinal}
r^{(1)}_{11}=0 \;\;{\mbox{for any}} \;\;s^{(1)}_{11}.
\end{eqnarray}

\begin{figure}
    \centering
    \includegraphics{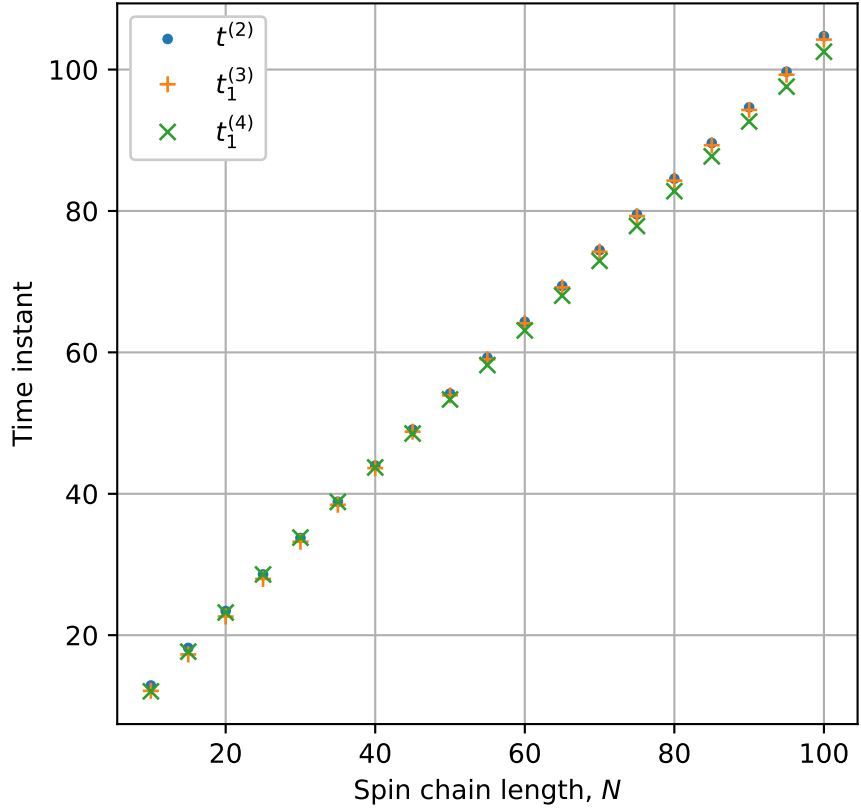}
    \caption{
        Time instants for  registration of a state transferred along the spin chain governed by $XX$-Hamiltonian (\ref{XX})
       in dependence on chain  length $N$.  
        The data correspond to Tables \ref{Table:time} and \ref{Table:time2}.
        Circle, crosses, pluses correspond to, respectively, 
        2-qubit sender including up to two excitations, 
        3- and 4-qubit sender including up to one excitation.
        Note, that transmission time slightly depend on the dimension of the sender.
    }
    \label{fig:time-instant-n}
\end{figure}

We fix the time instant for state registration as the time instant $t^{(N^{(S)})}_1$  maximizing the sum of probabilities of the excitation transfer from any node of the sender to any node of the receiver with $\varphi=0$, i.e. the maximum of the Frobenious norm of ${\cal{W}}$:
\begin{eqnarray}\label{max2}
\max_t || {\cal{W}}(t,0)||=  || {\cal{W}}(t^{(N^{(S)})},0)||.
\end{eqnarray}
For the 3- and 4-node sender these time instants $ t^{(3)}_1$ and $ t^{(4)}_1$  are given in Table \ref{Table:time2}  for different chain lengths, see also Fig.\ref{fig:time-instant-n}.
\begin{table}
\begin{tabular}{|c|cccccccccc|}\hline
 N & 10 & 15 & 20 & 25 & 30 & 35 & 40 & 45 & 50 & 55 \\\hline
 $t^{(3)}_1$&12.1286 & 17.2800 & 22.6386 & 27.9575 & 33.2243 & 38.4457 & 43.6308 & 48.7867 &
   53.9178 & 59.0261 \\\hline
$t^{(4)}_1$&
  12.0631& 17.6689&23.2101&28.5882&33.8054&38.8500& 43.7263& 48.5206&53.3406&58.2072\\\hline
\end{tabular}
\begin{tabular}{|c|ccccccccc|}\hline
 $N$ & 60& 65 & 70 & 75 & 80 & 85 & 90 & 95 & 100 \\\hline
$t^{(3)}_1$& 64.1135 & 69.1819 & 74.2338 & 79.2704 & 84.2923 & 89.2997 & 94.2938 & 99.2753 &
   104.245 \\\hline
$t^{(4)}_1$&
 63.1042& 68.0181& 72.9414& 77.8690&82.7979& 87.7267&  92.6558&97.5861&102.5179
 \\\hline
\end{tabular}
\caption{Time instants for state registration for spin chain of different chain length $N$ governed by $XX$-Hamiltonian (\ref{XX}) with  3- and {4-qubit sender including  up to one excitation.}}
\label{Table:time2}
\end{table}

{ The system (\ref{rsred21})-(\ref{rsred4}) must be replaced with the following one:}
\begin{eqnarray}\label{rsred2}\label{rsred1}
&&r^{(1)}_{i,j} = s^{(1)}_{i,j}, \;\; 2\le i,j \le N^{(S)} 
,\\\nonumber
&& s^{(1)}_{i, 1} =0, \;\; 2\le i \le N^{(S)},\\\nonumber
&&r^{(1)}_{1 ,1} =s^{(0)}.
\end{eqnarray}
In this case the general form of the block $s^{(1)}$ of the  initial density matrix of the $N^{(S)}$-qubit~sender
is~(\ref{tssK})~and~(\ref{ts}) with~${K=1}$.
Unitary transformation~$U$~(\ref{U}) includes the only nontrivial block~$U^{(1)}$ with parameters~$\varphi=\varphi^{(1)}$,
\begin{eqnarray}\label{U1}
U(\varphi)={\mbox{diag}}(1,U^{(1)}(\varphi)).
\end{eqnarray}

\subsubsection{Numerical optimization}
Since the solution space of the system (\ref{T1}), (\ref{T12}) is larger than the solution space of the reduces system (\ref{TW}), we use the former in the numerical optimization protocol.

The construction of the matrix that can be perfectly transferred consists of the three steps.
\begin{eqnarray}\label{Steps}
{\mbox{{\it Step 1:}}}&&
{\mbox{Solve system~(\ref{T1}), (\ref{T12})}}
\\\nonumber&&
{\mbox{to fix the parameters~$\varphi$ of the unitary transformation~(\ref{U1})}}\\\nonumber
{\mbox{{\it Step 2:}}}&&
{\mbox{Solve system~(\ref{rsred2})}}\\\nonumber&&
  {\mbox{for the elements of the initial density matrix of the sender}}\\\nonumber
{\mbox{{\it Step 3:}}}
&&
{\mbox{Perform the unitary transformation
  $U^{ex}$}}
  \\\nonumber 
&&{\mbox{exchanging positions of $r^{(0)}$ and $r^{(1)}_{1, 1}$}}
  \end{eqnarray}

{
The solution $\varphi=\varphi_0$ of the system~(\ref{T1}), (\ref{T12}) obtained at the first step provides the required structure of $r^{(1)}(t_0,\varphi_0)$, but this solution  is not unique because the number of $\varphi$-parameters is, generally, lager then the number of equations in the system (\ref{T1}), \ref{T12}). Therefore 
we can select a desired one. Namely,
we are interested in such \textit{optimal} $\varphi=\varphi^\mathrm{opt}$ parameters of the unitary transformation~(\ref{U1})
that provide the PTZ state obtained as a solution~of~(\ref{rsred2})
with the maximal deviation~$\delta^\mathrm{max} = \delta \vert_{\varphi=\varphi^\mathrm{opt}}$, where
\begin{eqnarray}\label{delF}
\delta(\varphi) =|| \rho^{(S)}_\infty -  \rho^{(R)}(\varphi) ||,
\end{eqnarray}
compare with  Eq.~(\ref{del}).
In Fig.\ref{fig:deviation-by-n}, $\delta^\mathrm{max}$ is given
as a function of the chain length~$N$ for 
$N^{(S)}=3,\;4$ and different lengths of the extended receiver~$N^{(ER)}$. {This figure demonstrates that $\delta^{max}$ increases with an increase in the dimension of the extended receiver and slowly decreases with length of the chain, see inset in the right-low corner of Fig.\ref{fig:deviation-by-n-s3}. Fig.\ref{fig:deviation-by-n-s3} demonstrates that the proposed optimization protocol works in the whole range of the considered chain length (up to 100) for the 3-qubit receiver (sender). Regarding the case of 4-qubit receiver (sender), the optimization protocol works in the whole range of $N$ only for large enough extended receiver, while the protocol with the minimal applicable dimension of the extended receiver ($N^{(A)} =1,2$) yields the satisfactory results only for $N\lesssim 50$ ($N^{(A)}=1$) or $N\lesssim 80$ ($N^{(A)}=2$), as  shown in Fig.\ref{fig:deviation-by-n-s4}. This fact just indicates that including more optimization parameters simplifies the search for the local minima.} We emphasize that in all cases the optimization leads to $r^{(1)}_{1,1}=0$ pointing that the system (\ref{TW}) yields the optimal parameters $\varphi$. 

\begin{figure*}[!]
  \centering
  \begin{subfigure}[c]{0.54\textwidth}
    \centering
    \includegraphics[width=\textwidth]{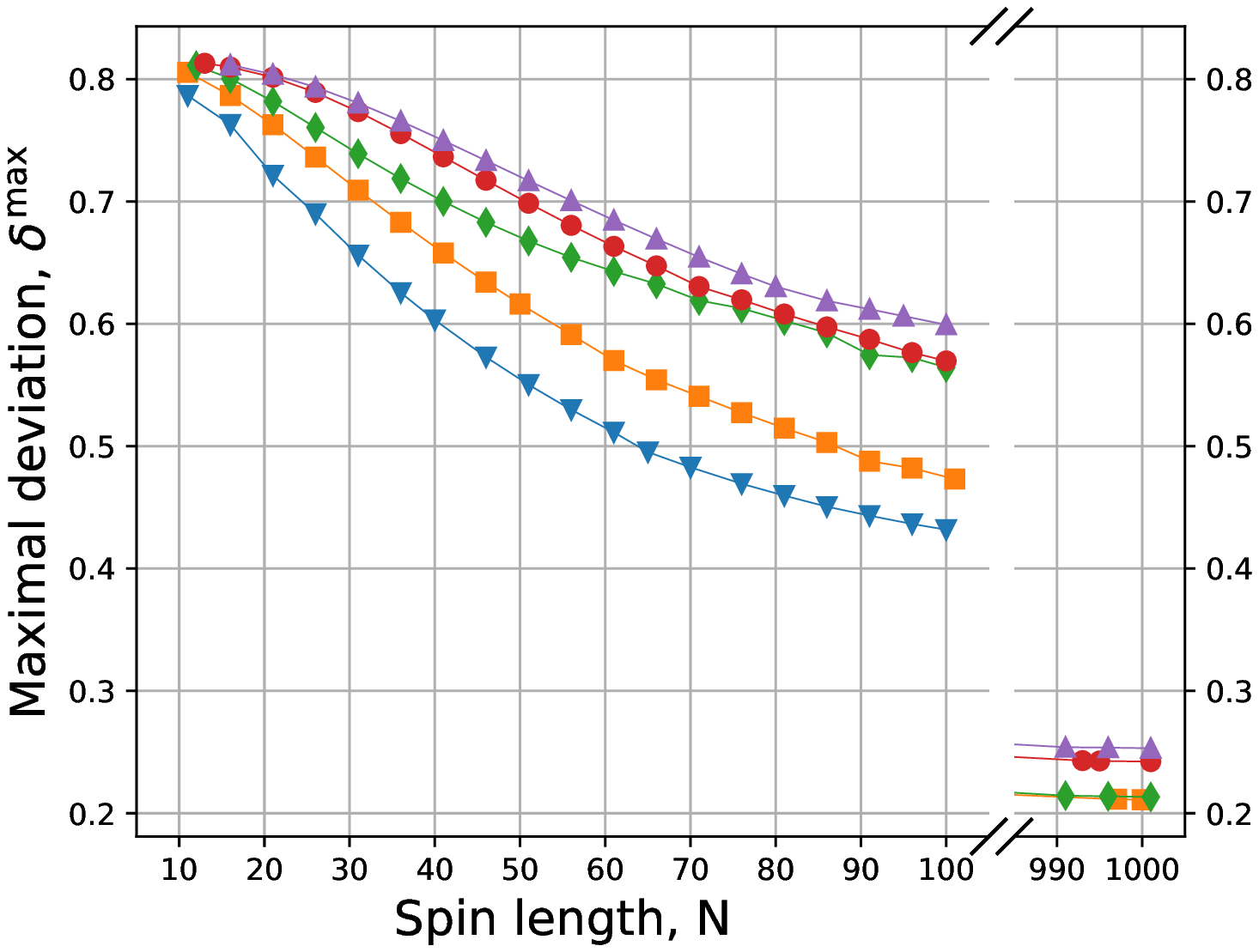}
    \caption{}
    \label{fig:deviation-by-n-s3}
  \end{subfigure}
  \hfill
  \begin{subfigure}[c]{0.42\textwidth}
    \centering
    \includegraphics[width=\textwidth]{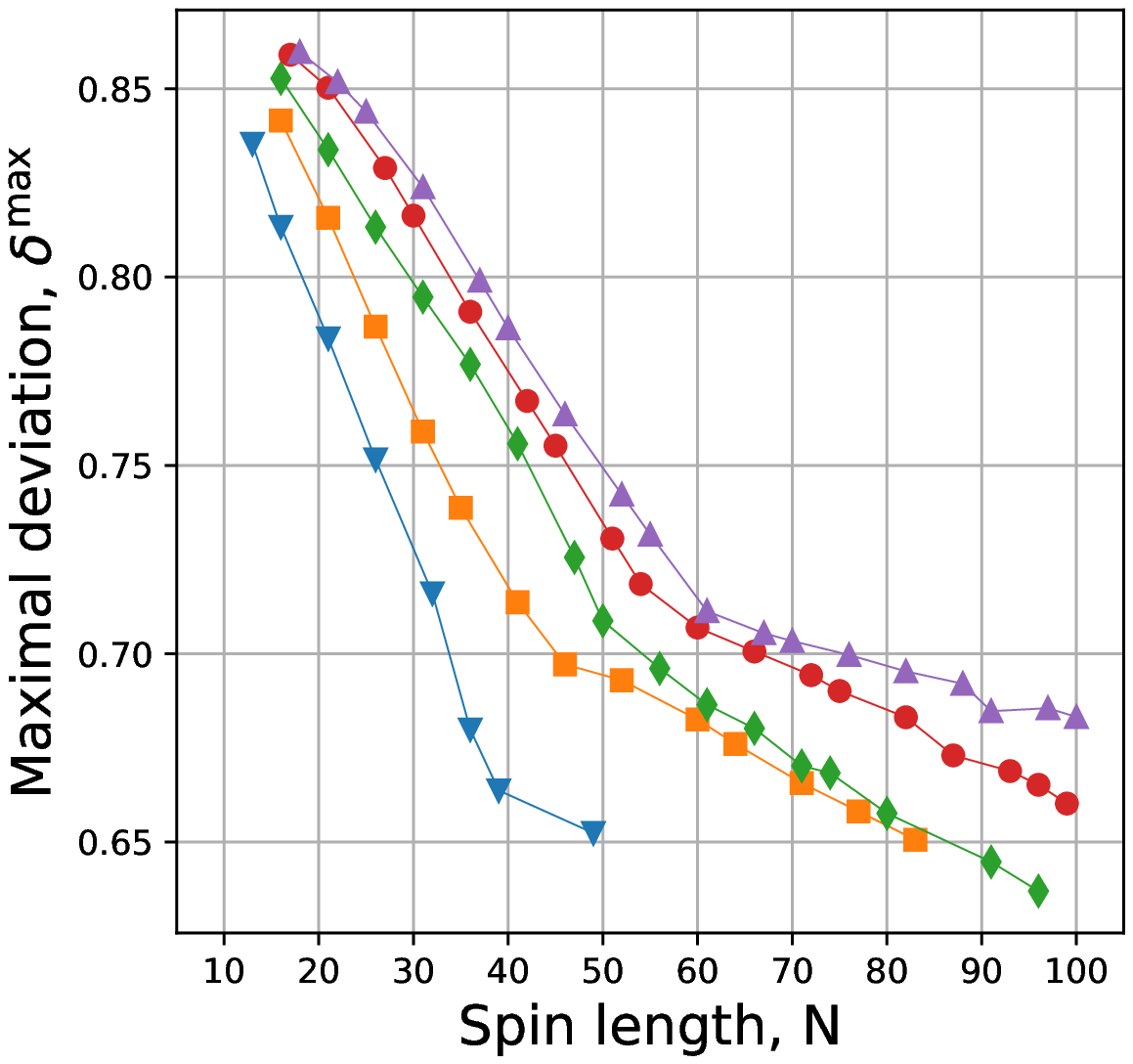}
    \caption{}
    \label{fig:deviation-by-n-s4}
  \end{subfigure}
  \caption{
    Deviation $\delta^\mathrm{max}$ of PTZ $\rho^{(S)}$
    from the asymptotic PTZ $\rho^{(S)}_{\infty}$ as a function of $N$.
    Optimization is performed for each spin length $N$ 
    and fixed number of ancillary spins $N^{(A)}$ in the extended receiver with size $N^{(ER)}=N^{(S)} + N^{(A)}$.
    Triangles, rhombus, circles, squares,inverted triangles correspond to, respectively,  $N^{(A)}=5$, $N^{(A)}=4$,   $N^{(A)}=3$,   $N^{(A)}=2$,   $N^{(A)}=1$.  {(a) $N^{(S)}=3$, the long-chain values of $\delta^{max}$ are shown at $N\sim 1000$ in the right-low corner of the plot; (b) 
     $N^{(S)}=4$.} 
  }
  \label{fig:deviation-by-n}
\end{figure*}

We notice that determining the optimal $\varphi^\mathrm{opt}$-parameters is not a trivial problem.
{The matter is that when constructing the target function we have to take into account the two-fold purpose of the optimization protocol. First, we have to satisfy the system  (\ref{T1}), (\ref{T12}).
Second, we have to maximize the deviation $\delta$. Therefore, the natural target function is
\begin{eqnarray}\label{TF}\label{eq:optimization-task}
F_T(\varphi) =w_1 S_T(\varphi)-w_2\delta(\varphi)
\end{eqnarray}
where
\begin{eqnarray}\label{ST}
S_T(\varphi)=\sum_{l,n,m} \Big|T^{(1)}_{l,1;n,m}(t^{(N^{(S)})}_1,\varphi )\Big| 
\end{eqnarray}
{(sum of all equations (\ref{T1}) and (\ref{T12}))}, 
 $w_i$, $i=1,2$ are some weights which are fixed below,  and sum is over all allowed values of the indexes $l$, $n$ and $m$, as in Eqs.(\ref{T1}), (\ref{T12}).
Here $\varphi$ is rather long set of parameters which varies from $12$ (4-qubit extended receiver) to 55 (8-qubit extended receiver) parameters. Thus, we combine Step 1 and Step 2 in the 3-step protocol (\ref{Steps}).

Let us describe the optimization protocol in more details. 
First, we replace $S_T$ given by (\ref{ST}) with the following one
\begin{eqnarray}\label{ST2}
S_T(\varphi)=\max_{l,n,m}  \Big|T^{(1)}_{l,1;n,m}(t^{(N^{(S)})}_1,\varphi )\Big|,
\end{eqnarray}
i.e. we define the residual $S_T$ as the maximal absolute value of the left sides of 
eqs. (\ref{T1}), (\ref{T12}). This residual removes  confrontation among different terms in
sum (\ref{ST}) extracting the most important one. 
}
}

What follows is based on two remarks.

{\bf Remark 1.} We study  optimization task~(\ref{eq:optimization-task}), (\ref{ST2}) with
a differential evolution (DE) algorithm~\cite{DifferentialEvolution1997,Wormington1999,Lampinen2002}
which is a kind of a genetic algorithm with crossover and mutation operations.
DE is a popular algorithm for a multiparameter optimization problems, including in quantum control~\cite{DEReview2020, PR2006,VolkovJPA2021}.
We perform calculation with the SciPy~package~\cite{SciPy} of  version~1.4.1.
The number of individuals in each population is $15 N^{(S)}$.
The probability of crossover is $CR = 0.7$ and probability of mutation randomly varies in the range $F \in [0.5, 1]$.
Some experiments were performed with higher population size~$1000  N^{(S)}$, higher mutation~$F=1.9$,
and lower recombination~$CR=0.3$ values to ensure finding the global minimum.

 {\bf Remark 2.}
We set in the functional~(\ref{eq:optimization-task}) weights as $w_1=w_2=1$,
because the natural ratio between $S_{T}$ and $\delta$ appearing during optimization  is suitable. For instance,  for some $\varphi=\varphi^{(approx)}_0$ we find that the residual is $S_{T}(\varphi^{(approx)}_0) \approx 10^{-3}$, while the deviation is $\delta(\varphi^{(approx)}_0) \approx 1$. Both these values are quite reasonable and convergence rates for both of them  are admittable. On the contrary, using larger ratio $(w_1S_t)/(w_2\delta)$ obtained  for some other weights in the formula (\ref{TF}) leads to a low convergence rate for $S_{T}$ while using smaller ratio  decreases the convergence rate for $\delta$.

{Thus, when minimizing $F_T(\varphi)$ with $S_T$ from (\ref{ST2}) we 
find an approximation~$\varphi=\varphi^\mathrm{approx}_0$}.
Now we polish the result with a local optimization method using $\varphi^{(approx)}_0$ for seed values of the parameters $\varphi$ and using  the same target function $F_T$ given by (\ref{TF}). 
The set of resulting parameters of unitary transformation we denote as ~$\varphi^\mathrm{approx}$.

Now we obtain the exact solution~$\varphi^\mathrm{opt}$ of  system~(\ref{T1}), (\ref{T12})  in the vicinity of~$\varphi^\mathrm{approx}$. This is the last step of the optimization protocol.

Step 3 of protocol (\ref{Steps}) can be done directly using the unitary transformation $U^{ex}$ exchanging the positions of $r^{(0)}$ and $r^{(1)}_{1,1}$ in the receiver state. 

Notice that increasing the dimension of the extended receiver leads to increasing the dimension of the optimization parameter space.
As shown in \cite{PT_2012,PI_2012}, in this case the number of traps, i.e. local but not global minima of the objective, generally decreases.
Thus the control landscape of the considered optimization problem for an  $n$-spin extended received is expected
to have less traps compared to the control landscape for $m$-spin  {extended receiver if} $n>m$.

To check the reliability of the global minimization 
we perform the optimization algorithm using two other global optimization methods: 
the  ``brute force'' method and  Dual Annealing method~\cite{Xiang1997} from SciPy~package~\cite{SciPy} of the version~1.4.1. 
In all three cases, the results are the same up to the absolute error of $10^{-3}$.


{
\subsection{Arbitrary parameter transfer via  0-order coherence matrix}
\label{Section:arb}
Let
the unitary transformation of the extended receiver be such that the operator ${\cal{W}}$ is diagonal, i.e. 
\begin{eqnarray}\label{WWW}
{\cal{W}}_{ij} =0,\;\;i\neq j, \;\;i,j=1,\dots,N^{(S)} .
\end{eqnarray}
System (\ref{WWW}) includes $\frac{1}{2}N^{(S)}(N^{(S)}-1)$ complex equations. 
In this case 
\begin{eqnarray}
&&T^{(1)}_{ij;nm}=\lambda_{ij} \delta_{in}\delta_{jm},\;\;\lambda_{ij}=(\lambda_{ji})^* \;\; , \;\;i,j,n,m=1,\dots,N^{(S)}\\\label{lam}
&&\lambda_{ij} = {\cal{W}}_{ii}  {\cal{W}}_{jj}^* .
\end{eqnarray}
Therefore any element in the receiver's one-excitation block is proportional to the appropriate element of the sender's one-excitation block, i.e.
\begin{eqnarray}\label{a}
&& r^{(1)}_{i,j}=\lambda_{ij}  s^{(1)}_{i,j},\;\;  i,j = 1,\dots,N^{(S)},
\end{eqnarray}
where $\lambda_{ij}$ are  scale factors, $|\lambda_{ij}|\le 1$.
Formula (\ref{a}) means the structural restoring of the whole one-excitation block of the 0-order coherence matrix \cite{Z_2018,FPZ_Arxiv2021}.
In this case $r^{(0)}$ (the only element of the zero-excitation block) provides the normalization and therefore can not be proportional to $s^{(0)}$. Unlike in Sec.~\ref{Section:01p}, we do not exchange positions of any diagonal elements and consequently all elements of the one-excitation block can be non-zero. 

Below we consider the cases of 2- and  3-qubit senders and appropriate extended receivers with unitary transformations whose parameters not only solve system (\ref{WWW}), but also maximize the scale factors $\lambda_{ij}$ in (\ref{a}). 
Therefore, we consider the minimal of $|\lambda_{ij}|$,
\begin{eqnarray}\label{Wnorm}
\lambda_{opt}=
\min_{i,j} |\lambda_{ij}|,
\end{eqnarray}
as a maximization object and  use the 
global minimization algorithm with the following target function:
\begin{eqnarray}
F_T= \sqrt{\sum_{i\neq j} |{\cal{W}}_{ij}|^2} - \min_{i,j} |\lambda_{ij}|.
\end{eqnarray}
At that, the optimization parameters are the parameters of the unitary transformation of the extended receiver $\varphi$ and time $t$.

However, to simplify calculation, we perform global optimization  over the parameters $\varphi$ at fixed time instants and find  the values  of $|\lambda_{opt}|$ inside of the time interval around $t\sim N$ as shown in Fig.~\ref{fig:ArbParTime} for 2-qubit receiver and 3-qubit extended receiver.
\begin{figure*}[!]
    \includegraphics[width=0.5\textwidth]{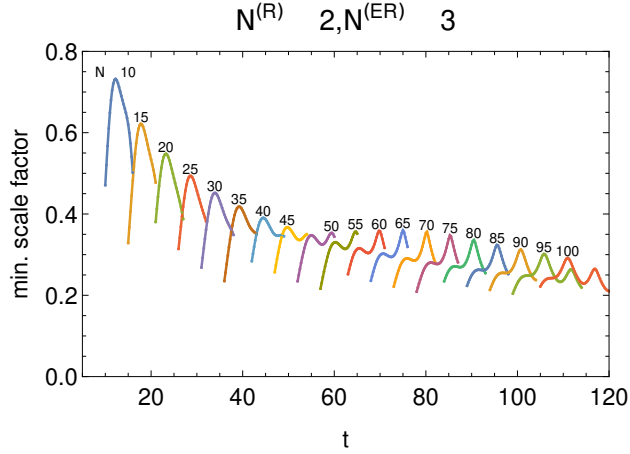}
    \caption{$N^{(S)}=2$, $N^{(ER)}=3$. Dependence of $|\lambda_{opt}|$ (minimal scale factor)  on time for different chain lengths from $N=10$ to $N=100$. Only time intervals around the assumed maxima are considered for each $N$.} 
    \label{fig:ArbParTime}
  \end{figure*}
We see in this figure that there are two local maxima near the optimal time instant. For $N<50$ the first local maximum is higher, while the second becomes higher for $N\ge 50$. This leads to a jump in the function $t_{opt}(N)$ and to a small local minimum in the function $|\lambda_{opt}(N)|$ as shown in Fig.~\ref{fig:TimeF} for the case $N^{(S)}=2$ and  $N^{(ER)}=3$.  This jump leads to the local minimum  on the graph of $|\lambda_{opt}|$  in Fig.~\ref{fig:FN}. Similar jump appears for the case  $N^{(ER)}=4$, although the graph of $|\lambda_{opt}|$ is monotonic.  In the case $N^{(S)}=3$, $N^{(ER)}=5$, $|\lambda_{opt}|$ has  three  local maxima near the optimal time instant, therefore two jumps appear on the graph in Fig.~\ref{fig:TimeF}. However, unlike the case $N^{(S)}=2$, the graph of $|\lambda_{opt}|$ remains monotonic. 
The choice of dimensions of extended receivers is explored in Appendix.

\begin{figure*}[!]
  \centering
  \begin{subfigure}[c]{0.48\textwidth}
    \centering
    \includegraphics[width=\textwidth]{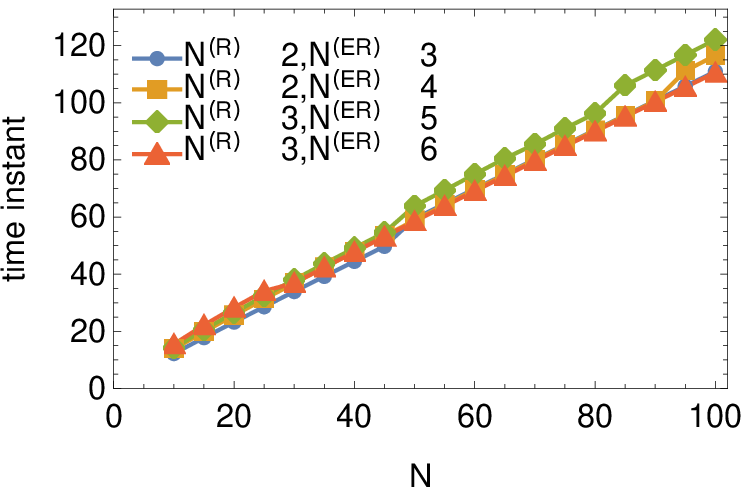}
    \caption{}
    \label{fig:TimeN}
  \end{subfigure}
  \begin{subfigure}[c]{0.48\textwidth}
    \centering
    \includegraphics[width=\textwidth]{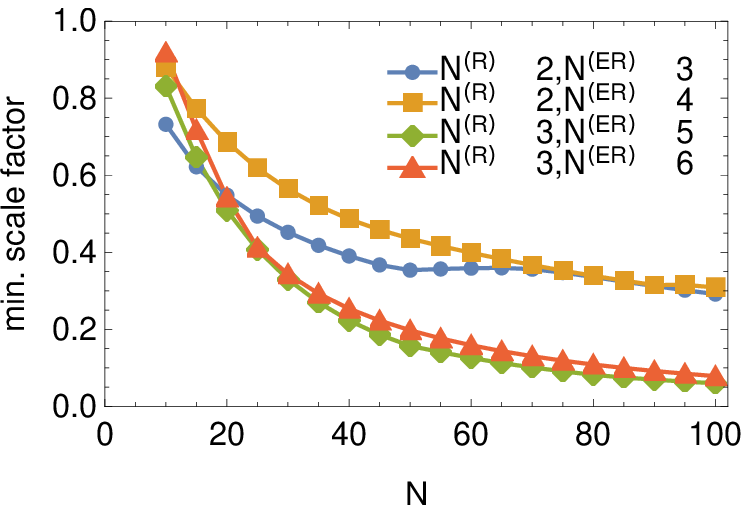}
    \caption{}
    \label{fig:FN}
  \end{subfigure}
  \caption{
   Time instant $t_{opt}$ (a) and  absolute value of the minimal scale factor $|\lambda_{opt}|$ (b) as functions of $N$ for 
   $N^{(S)}=2$, $N^{(ER)}=3$, $N^{(S)}=2$, $N^{(ER)}=4$ and $N^{(S)}=3$, $N^{(ER)}=5$.
  }
  \label{fig:TimeF}
\end{figure*}
}

\section{Conclusions}
\label{Section:conclusion}
Unlike the elements of the higher-order coherence matrices,  not all  diagonal elements of the 0-order coherence matrix  in general can be structurally restored~\cite{FPZ_Arxiv2021}. This happens due to the trace-normalization condition which  0-order coherence matrix must satisfy.   However, the normalization condition  creates a possibility for the perfect transfer of 0-order coherence matrix with elements fixed in a certain way. We show that the perfect transfer can be produced using  one additional unitary transformation applied to the receiver at the time instant of state registration to exchange positions of two diagonal elements. One of these elements is necessarily the only element of the 0-excitation block and another one is, in the simplest case,  another one-element block of the receiver density matrix, associated with  $N^{(S)}$-excitation subspaces of the $N^{(S)}$-qubit receiver. If this element does not exist (this happens if the number of excitations appeared in the receiver state is less then $N^{(S)}$) then any other diagonal element can be taken. 
This is the case when we need to use the unitary transformation of the extended receiver to provide consistency of the protocol. 

We show that, with an increase in the chain length,  the perfectly transferable sender density matrix tends to the diagonal matrix $\rho^{(S)}_\infty$ with only one non-zero element.

We study the deviation $\delta$ of the 0-order coherence matrix  perfectly transferred through the $N$-qubit chain  from the limiting matrix $\rho^{(S)}_\infty$ for a particular case of transferred states with 0- and 1-excitation blocks of 3- and 4-qubit receiver (and sender) using the  ($N^{(S)}+i$)-qubit  extended receiver with $i=1,2,3,4$, thus verifying  that  increasing the dimension of the extended receiver we can increase the deviation $\delta$.  

For the practical purpose
the presence of  deviation from the trivial asymptotic is important, because positivity of the density matrix requires vanishing of all elements from the row and column where the zero diagonal element  appears. Therefore, the asymptotic 0-order coherence matrix  $\rho^{(S)}_\infty$ can not transfer any other element.

{We also remark, that the simplest case of perfectly transferred  1-excitation 0-order coherence matrix can not be used to transfer the higher order coherence matrices (1-order in this case) because these elements belong to the first row and first column of the receiver's density matrix which are zeros according to Eq.(\ref{BlocksS3}).}

The 0-order coherence matrix of the receiver state includes one more feature. {The  elements of the 1-excitation block of this matrix can be restored similarly to the elements of the higher-order coherence matrices. At that, however, the 0-excitation element provides the trace-normalization. This fact allows to use the elements of this block  to transfer arbitrary parameters from the sender to the receiver. Of course, there is no perfect transfer of 0-order coherence matrix in this case. We study the absolute value of the minimal of the scale factors  ahead of the transferred arbitrary parameters as a function of the chain length for the case of 2- and 3-qubit receiver with  the appropriate minimal dimension of the  extended receivers. }

We have to emphasize that the perfect transfer of the 0-order coherence matrix implies certain constraints on the structure of the restorable higher-order coherence matrices. Namely, the rows and columns of the transferred matrix corresponding to the diagonal elements which are exchanged by the final unitary transformation of the receiver must be zero, {see Eq.(\ref{blocksS2}).} 

{We also shall emphasize that the unitary transformation constructed in Sec.~\ref{Section:01p} provides the  PTZ for a particular 0-order coherence matrix, while the unitary transformation constructed in Sec.~\ref{Section:arb} for transferring the  arbitrary parameters  is universal in the sense that the same transformation can be used to transfer parameters of any initial sender's state. The scale factors appearing ahead of the arbitrary parameters in the receiver's state are permanent characteristics of the protocol similar to~\cite{FPZ_Arxiv2021}.}

To resume, we have studied the structure of the 0-order coherence matrix which can be perfectly transferred along the spin chain. 
At that the  elements of the 1-excitation block of the 0-order coherence matrix can be also used to transfer arbitrary parameters if the unitary transformation of the extended receiver is properly adjusted. Of course, all the transferable arbitrary parameters must keep the positivity of the associated density matrices.    

{We also emphasize that the proposed protocol of the perfect transfer of the 0-order coherence matrix is not related to just the $XX$ Hamiltonian (\ref{XX}). It can use any Hamiltonian satisfying commutation condition (\ref{comm}), for instance, $XXZ$ Hamiltonian.  Although we consider the nearest neighbor  interaction, it is not a necessary requirement to the Hamiltonian and we can include the remote node interactions (e.g. dipole-dipole interactions) as far as this step doesn't destroys the above commutation condition. Also the homogeneous chain can be replaced with any non-homogeneous 
chain since the Hamiltonian for a non-homogeneous chain  also satisfies 
the commutation condition (\ref{comm}). Of course, all the above  modifications of the spin system (changing the Hamiltonian, including  the remote node interaction and  passing to the non-homogeneous chain) change the required unitary transformation $U$ (\ref{U})  and  the form of the perfectly transfered 0-order coherence matrix $\rho^{(S)}$ derived in Sec.\ref{Section:01}, but the described protocol of  PTZ is stable to all those modifications.}

This work was funded by Russian Federation represented by the Ministry of
Science and Higher Education (grant number 075-15-2020-788).

\section{Appendix: Representation of transfer operator $T$ in terms of evolution operator and unitary transformation of extended receiver }
\label{Section:A}

According to Ref.~\cite{FPZ_Arxiv2021}, the receiver density matrix is defined by two unitary operators. The first one, denoted by $V$, describes the evolution under certain Hamiltonian $H$: $V(t)=e^{-i Ht}$. This operator acts on the whole system. The second operator $U$ acts only on the extended receiver, it depends on the set of free parameters which are used to satisfy the requirements of the target state creation. These two operators can be combined into the single operator $W$ (\ref{WW}). Therefore, the receiver's density matrix (\ref{blocksR})  reads
\begin{eqnarray}\label{WWR}
\rho^{(R)}={\mbox{Tr}}_{S,TL}\left(W(t,\phi) \Big( \rho^{(S)}(0)\otimes \rho^{(TL,R)}(0)\Big) W^\dagger(t,\varphi)\right).
\end{eqnarray}

Now we introduce multiindexes associated with the sender ($S$), transmission line ($TL$) and receiver ($R$) 
\cite{FPZ_Arxiv2021}. These indexes are represented by the capital latin letters with appropriate subscript $S$, $TL$ ar $R$. Then we write  receiver density matrix (\ref{WWR}) as
\begin{eqnarray}
r^{(k)}_{N_R,M_R}& =& \sum_{N_S,N_{TL}} W^{(k)}_{N_S,N_{TL},N_R;I_S,0_{TL},0_{R}} \rho^{(S)}_{I_S,J_S}
(W^{(k)})^\dagger_{J_{S},0_{TL},0_R;N_S,N_{TL},M_R},\\\nonumber
&&k=0,\dots, N^{(S)}.
\end{eqnarray}
In the 0- and 1-excitation case, for the elements of the 0-order coherence matrix we have
\begin{eqnarray}\label{0ex}
&&\rho^{(R)}_{0_R,0_R}\equiv r^{(0)} =1-\sum_{i=1}^{(N^{(S)}} r^{(1)}_{ii},\\\label{1ex} 
&&r^{1}_{N_R,M_R} =  W^{(1)}_{0_S,0_{TL},N_R;I_S,0_{TL},0_{R}} s^{(1)}_{I_S,J_S}
(W^{(1)})^\dagger_{J_{S},0_{TL},0_R;0_S,0_{TL},M_R}.
\end{eqnarray}
We can introduce $N^{(S)}\times N^{(S)}$ matrix  ${\cal{W}}$ passing from the multiindex basis to the computational basis following the rule 
\begin{eqnarray}
(\underbrace{0\dots 0}_{i-1} 1 \underbrace{0\dots 0}_{N^{(S)}-i}) \to i ,
\end{eqnarray}
so that 
\begin{eqnarray}
{\cal{W}}_{ij}=
W^{(1)}_{0_S,0_{TL},\underbrace{0\dots 0}_{i-1} 1 \underbrace{0\dots 0}_{N^{(S)}-i};\underbrace{0\dots 0}_{j-1} 1 \underbrace{0\dots 0}_{N^{(S)}-j},0_{TL},0_{R}}.
\end{eqnarray}
and (\ref{1ex}) gets the matrix form (\ref{calW}).

Let us estimate the minimal dimension of the extended receiver required to satisfy conditions   (\ref{TW}).
The number of  parameters in the unitary transformation is defined  by Eq.~(\ref{utr}):
$N^{(ER)} (N^{(ER)}-1)$. {However, according to Eq.(\ref{1ex}) and  definition of $W$ (\ref{WW}), only $N^{(S)}$ rows  of the unitary transformation $U$ are included into ${\cal{W}}$. Taking into account the normalization of rows of $U$ and disregarding the common phase in each row, the number of effective free real parameters is
\begin{eqnarray}
\sum_{j=1}^{N^{(S)}}  (2 N^{(ER)} - 2 j) = N^{(S)}(2N^{(ER)}-N^{(S)}-1).
\end{eqnarray}
This number is not less then  the number of scalar real equations in (\ref{TW}) (which is $2 N^{(S)}$) for $N^{(ER)}\ge \frac{3+N^{(S)}}{2}$. However, $N^{(ER)}\le  N^{(S)}$ means that the extended receiver is not bigger then the receiver,
i.e., the unitary transformation $U$ is applied to the receiver state.} Then we can write 
\begin{eqnarray}
{\mbox{rank}} \left(\rho^{(S)}(0)\right)& =& {\mbox{rank}} \left({\mbox{Tr}}_{TL,S}  V(t) \rho^{(S)}(0) V^\dagger (t)\right)
\nonumber\\
&{>}&
{\mbox{rank}} \left( U  {\mbox{Tr}}_{TL,S} V(t) \rho^{(S)}(0) V^\dagger (t) U^\dagger\right) =
{\mbox{rank}} \left(\rho^{(R)}(t)\right).
\end{eqnarray}
But the unitary transformation $U$ can not reduce the rank of the matrix. Therefore, $N^{(ER)}\ge  N^{(S)}+1$. This inequality is confirmed by the numerical calculations for 3- and 4-qubit receiver in Sec.~\ref{Section:01p}, Fig.~\ref{fig:deviation-by-n}. 

Now we estimate the minimal dimension of the extended receiver needed to optimize the solution of system 
(\ref{WWW}).
By construction, the $N^{(S)}\times N^{(S)}$ matrix ${\cal{W}}$ is a product of two matrices 
\begin{eqnarray}
{\cal{W}}=\tilde U \tilde V,
\end{eqnarray}
where $\tilde U$ is a $N^{(S)}\times N^{(ER)}$ matrix of last $N^{(S)}$ rows of the unitary transformation of the extended receiver and 
$\tilde V$ is the $N^{(ER)}\times N^{(S)}$ matrix which is left-down corner block of the evolution matrix $V$.
Let $U$ be composed of the rows $a_i$, $i=1,\dots,N^{(ER)}$, $\tilde U$ be composed of the rows $a_i$, $i=1,\dots,N^{(S)}$ and $\tilde V$ be composed of the columns $b_i^\dagger$, $i=1,\dots,N^{(S)}$. Then the
diagonal form of ${\cal{W}}$ requires
\begin{eqnarray}\label{ab0}
a_i b_j^\dagger =0,\;\;i\neq j, \;\;i,j=1,\dots,N^{(S)}.
\end{eqnarray}
{Then we can expand $b_i$ in the basis of $a_i$ as follows
\begin{eqnarray}\label{bb}
b_i=\alpha_{ii} a_{i} +\sum_{j=N^{(S)}+1}^{N^{(ER)}} \alpha_{ij} a_j,\;\;i=1,\dots,N^{(S)}
\end{eqnarray}
where $\alpha_{ij}$ are some constant coefficients. 

Now we show that ${\cal{W}}$ can be diagonalized only if $N^{(ER)}\ge 2N^{(S)}-1$. In fact, suppose that the diagonalization can be done for the case 
$N^{(ER)}<  2 N^{(S)}-1$ and the system (\ref{bb}) is obtained. We solve  first  $N^{(ER)}- N^{(S)}$ equations of system (\ref{bb}) for $a_j$, $j>N^{(S)}$:
\begin{eqnarray}
a_j=\sum_{i=1}^{N^{(ER)}- N^{(S)}} \gamma_{ji} (b_i -\alpha_{ii} a_i).
\end{eqnarray}
where the coefficients $\gamma$'s depend on $\alpha$'s.
Then the rest $N^{(S)} - (N^{(ER)}- N^{(S)}) =2 N^{(S)} -N^{(ER)}$ vectors $b_i$ we have
\begin{eqnarray}
b_k &=& \alpha_{kk} a_k + \sum_{j=N^{(S)}+1}^{N^{(ER)}} \alpha_{kj} \sum_{n=1}^{N^{(ER)}- N^{(S)}} \gamma_{jn} (b_n -\alpha_{nn} a_n),\\\nonumber
k&=& N^{(ER)}- N^{(S)}+1 , \dots , N^{(S)}.
\end{eqnarray}
or
\begin{eqnarray}
b_k  - 
\sum_{j=N^{(S)}+1}^{N^{(ER)}}  \sum_{n=1}^{N^{(ER)}- N^{(S)}}\alpha_{kj} \gamma_{jn} b_n &=& \alpha_{kk} a_k + \sum_{j=N^{(S)}+1}^{N^{(ER)}} \sum_{n=1}^{N^{(ER)}- N^{(S)}} \alpha_{kj} \gamma_{jn} \alpha_{nn} a_n,\\\nonumber
k&=& N^{(ER)}- N^{(S)}+1 , \dots , N^{(S)}.
\end{eqnarray}
By virtue of (\ref{ab0}), we can write
\begin{eqnarray}\label{bb2}
b_k b_j^+ =0,\;\; j,k= N^{(ER)}- N^{(S)}+1 , \dots , N^{(S)},\;\;j\neq k.
\end{eqnarray}
Notice that system (\ref{bb2})
represent  the additional relations between $b_i$.
However, the vectors $b_i$ are fixed by the Hamiltonian and they can  not satisfy these additional relations in general. Therefore,  we have to impose the requirement $N^{(ER)}\ge 2N^{(S)}-1$. According to this requirement, $N^{(ER)}\ge 3$ for $N^{(S)}=2$ and $N^{(ER)}\ge 5$ for $N^{(S)}=3$, which is used in Sec.~\ref{Section:arb}, see Figs.~\ref{fig:ArbParTime} and~\ref{fig:TimeF}.
}

{\bf Data availability statement.} The authors confirm that the data supporting the findings of this study are available within the article.

\end{document}